\documentclass[prl,twocolumn,showpacs,preprintnumbers,amsmath,amssymb]{revtex4}

\usepackage{graphicx}
\usepackage{epstopdf}
\usepackage{dcolumn}
\usepackage{bm}

\begin{document}


\title{Correlation of anomalous scattering and superconductivity in Pb$_{.99}$Tl$_{.01}$Te doped with additional indium donors}

\author{A. S. Erickson,$^{1,2}$ N. P. Breznay,$^{2}$  E. A. Nowadnick,$^{1,3}$ T. H. Geballe,$^{1,2}$ I. R. Fisher$^{1,2}$} 
\affiliation{1. Stanford Institute for Materials and Energy Sciences, SLAC National Accelerator Laboratory, 2575 Sand Hill Road, Menlo Park, CA 94025.\\2. Geballe Laboratory for Advanced Materials and Department of Applied Physics, Stanford University, CA 94305.\\3. Geballe Laboratory for Advanced Materials and Department of Physics, Stanford University, CA 94305.
}

\date{\today}

\begin{abstract}
Recent evidence for a charge-Kondo effect in superconducting samples of Pb$_{1-x}$Tl$_x$Te \cite{yana} has brought renewed attention to the possibility of negative U superconductivity in this material, associated with valence fluctuations on the Tl impurity sites \cite{joerg}.  Here, we use indium as an electron-donor to counterdope Pb$_{.99}$Tl$_{.01}$Te and study the effect of the changing chemical potential on the Kondo-like physics and on the superconducting critical temperature, $T_c$.  We find that, as the chemical potential moves away from the value where superconductivity, Kondo-like physics, and chemical potential pinning are expected, both $T_c$ and the low-temperature resistance anomaly are suppressed.  This provides further evidence that both the superconductivity and the Kondo-like behavior are induced by the same source, as anticipated in the negative U model.

\end{abstract}

\maketitle

When doped with more than $x_c =$ 0.3\% Tl atoms, the degenerate semiconductor, Pb$_{1-x}$Tl$_x$Te, superconducts \cite{yana} with $T_c$ reaching 1.4 K at the solubility limit, $x = 1.4\%$ \cite{murakami, nemov1998}, an order of magnitude higher than other materials of similarly low hole concentration, $n_H$ = 10$^{19}$cm$^{-3}$ \cite{hulm1969}. Thallium is the only element to induce superconductivity in this host, though it can be doped to similar carrier concentrations with Na \cite{dornhaus}.  For Tl concentrations where superconductivity is observed, a low temperature resistance minimum is also found, with a temperature-dependence consistent with the Kondo effect with a Kondo temperature, $T_K = 6$ K, though magnetic impurities were ruled out as a source \cite{yana}.  Significantly, Tl is a valence-skipping element, preferring $+3$ and $+1$ filled-shell valence states to the $6S^1$ configuration of the $+2$ valence state \cite{varma}. Measurements of the Hall coefficient have shown that, for low Tl concentrations, Tl enters in the $+1$ valence state, doping holes and changing the chemical potential. When $x$ exceeds the critical concentration for superconductivity, $x_c = 0.3 \%$, the Hall number stabilizes, indicating an average Tl valence of $+2$, and pinning of the chemical potential \cite{murakami, yanaprb}.  Due to the valence skipping nature of Tl atoms, a mixed-valence state is likely for $x > x_c$ \cite{drabkin1981}.  Preliminary core level photoemission spectra at  $x > x_c$ are consistent with this expectation \cite{kaminski}.

Mixed-valence impurities can be described by a Hubbard model with a negative value of the parameter, $U$, the charging energy for having two electrons on the same impurity site \cite{anderson}. In this model, the energy difference between the empty shell Tl$^{+3}$ and the filled-shell Tl$^{+}$ state depends on the chemical potential, $\mu$ \cite{joerg}.  If the energy difference between these two states is small, it may be possible to adjust the chemical potential through doping to the value, $\mu = \mu_c$, where the energy difference between these two states is zero.  Once $\mu = \mu_c$, further doping will yield a mixture of $+1$ and $+3$, pinning the chemical potential at this critical value \cite{drabkin1981}. In the model of negative U superconductivity, resonant 2-electron scattering between degenerate $+1$ and $+3$ valence states can serve as a mechanism of forming cooper pairs between valence-band holes, and also contributes Kondo-like behavior \cite{joerg, moizhes1981, taraphder1991}. This model has been shown to describe Pb$_{1-x}$Tl$_x$Te well \cite{joerg}. 

The correlation between chemical potential pinning, Kondo-like scattering, and superconductivity implied in the negative U model can be tested by changing the chemical potential by doping of additional holes or electrons in samples with Tl concentration, $x > x_c$. The correlation between chemical potential pinning and superconductivity was studied in the case of Pb$_{1-x-y}$Tl$_x$Li$_y$ \cite{kaidanov1987} and Pb$_{1-x-y}$Tl$_x$Na$_y$Te \cite{chernik1982, kaidanov1986}, where Li and Na counterdopants are used to inject additional holes, while keeping a fixed Tl concentration. In these cases, moving the chemical potential away from the critical value, $\mu_c$, resulted in a suppression of $T_c$ \cite{kaidanov1987,chernik1982}.  In the present paper, we study the effect of indium counter-dopant concentration on the electronic properties of Pb$_{.99-y}$Tl$_{.01}$In$_y$Te, for a fixed Tl concentration, $x = 1.0(3)\% > x_c = 0.3\%$, focusing on the correlation between Kondo-like scattering, which was not investigated in previous studies, and chemical potential pinning.

When doped into PbTe, In acts as an electron donor \cite{averkin, kaidanov1973}. We find that, similar to the case of Na and Li counterdopants, for which the dopants were electron acceptors, superconductivity is suppressed as the chemical potential, $\mu$, moves away from the critical value, $\mu_c$ at a critical In counterdopant concentration of $y_c = 1.2\%$.  This is accompanied by a change in the Hall coefficient, and a reduction in the density of states at the Fermi level, $N(0)$, derived from the electronic coefficient of specific heat, $\gamma$.  Significantly, we also find a suppression of the low temperature resistance minimum, and an increase in the residual resistance ratio, $RRR$, as $y$ increases beyond $y_c = 1.2\%$, demonstrating a strong correlation between superconductivity and the anomalous low-temeperature scattering previously attributed to the charge-Kondo effect \cite{yana}.

\begin{table}
\centering

\caption{ \label{comp} Composition of Pb$_{1-x-y}$Tl$_x$In$_y$Te samples used in this study, determined by EMPA measurements.  Error bars are standard deviations of 4-6 data points.}
\begin{tabular}{| c | c | c | c | c | c | c | c |}
\hline
$x$ (at. \%) & 1.1(2) & 1.0(4) & 0.7(4) & 1.0(2) & 1.0(4) & 1.0(4) & 1.1(4)\\ 
\hline
$y$ (at. \%) & 0 & 0.5(1)  & 0.8(1) & 1.2(1) & 1.4(1) & 2.1(1) & 3.1(1)\\ 
\hline
\end{tabular}

\end{table}

Single crystals of Pb$_{1-x-y}$Tl$_{x}$In$_y$Te were grown using the physical vapor tranpsort method described elsewhere \cite{yanaprb}.  The composition was measured by electron microprobe analysis (EMPA), using PbTe, In metal, and Tl$_2$Te standards.  The results of these measurements are shown in table \ref{comp}.  The large standard deviations found for the Tl impurity concentration in samples also containing In, relative to the standard deviation found for the In-free sample, likely reflect slight inhomogeneity in Tl concentration across any given sample. As each batch was grown independently, it was not possible to maintain exactly equal Tl concentrations throughout the series.  However, all samples were within one standard deviation of the average value $x = 1.0\%$, a Tl concentration of $x = 1.0(5)\%$ will be assumed in the following analysis.

\begin{figure}
\centering
\includegraphics[width=3.5in]{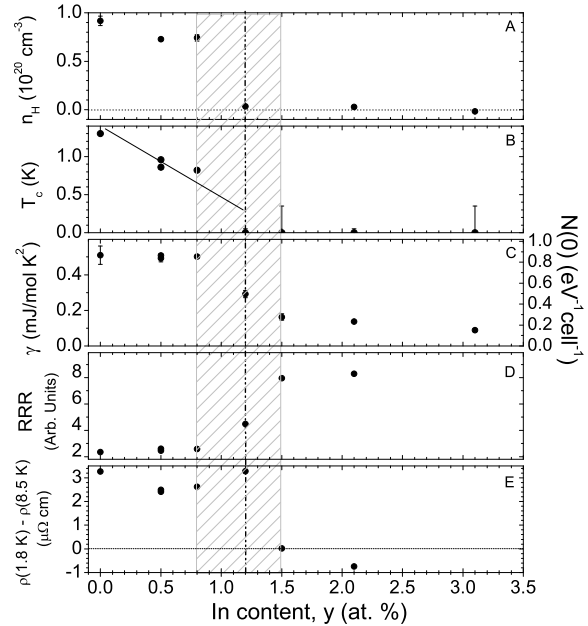}
\caption{\label{trendsintl} (A) The Hall number, $n_H$, in Pb$_{.99-y}$Tl$_{.01}$In$_y$Te as a function of In concentration, $y$.  (B)  Superconducting transition temperature, $T_c$, derived from heat capacity measurements.  The line is a guide to the eye.  (C) (left axis) electronic coefficient to the specific heat, $\gamma$, and (right axis) corresponding density of states per spin at the Fermi level, $N(0)$. (D) $RRR$, defined as R(300 K) / R(1.8 K).  (E) Resistance minimum depth, defined as $\rho$(1.8 K) - $\rho$(8.5 K).  The vertical dot-dash line represents the critical In concentration below which superconductivity and Kondo-like behavior are observed, $y_c = 1.2\%$, and the hashed region denotes the extent of crossover behavior.}
\end{figure}

\begin{figure}
\centering
\includegraphics[width=3.5in]{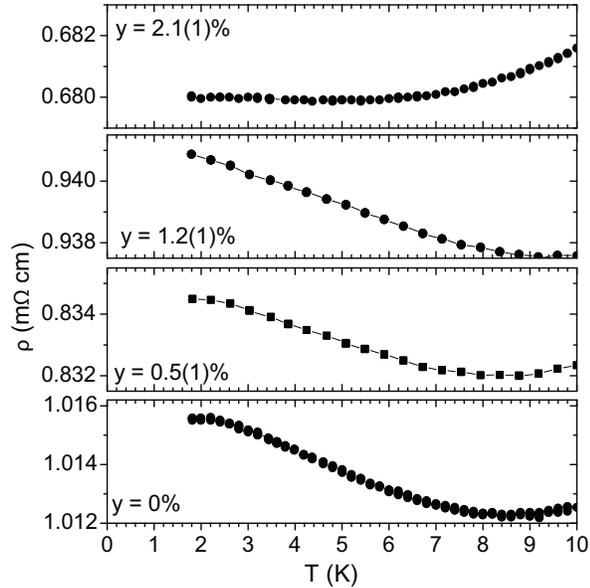}
\caption{\label{rholowtintl}Representative resistivity data showing behavior near the resistance minimum, for Pb$_{.99-y}$Tl$_{.01}$In$_{y}$Te, with $0\% < y < 2.1\%$.}
\end{figure}

Bars were cleaved from single crystals for electrical tranpsort measurements.  Resistivity data were obtained at frequencies of 13.5 Hz, 13.7 Hz or 37 Hz, and current densities along the [100] direction of order 100 mA/cm$^2$ at temperatures above 1.8 K and 10 mA/cm$^2$ for data taken below 1.8 K, using a Quantum Design PPMS system equipped with a He$^3$ insert. Measurements below 350 mK were performed using a He$^3$/He$^4$ dilution refrigerator. Data below 1.8 K were taken at a variety of current densities to check for heating effects.  The Hall number at a temperature of 1.8 K was obtained from linear fits to the transverse voltage in fields from -9 to 9 T, aligned along the [001] direction. All electrical contacts were made by sputtering gold contact pads, then fixing platinum wire to the pads using Epotek H20E conductive silver epoxy, with typical contact resistances between 2 and 5 $\Omega$.  

Although PbTe is a multi-band semiconductor, the Hall number,  $n_H = 1/R_He$, shown as a function of $y$ in panel (a) of figure \ref{trendsintl}, can be used as a reasonable estimate of the carrier concentration because the mobilities of the two valence bands occupied at these carrier concentrations are similar \cite{yanaprb}.  For In concentrations, $y < 0.9\%$, the carrier concentration is near $0.8 \times 10^{20}$ cm$^{-3}$, indicating the chemical potential is pinned at the critical value $\mu_c$, as observed in Pb$_{1-x}$Tl$_x$Te, with $x > 0.3\%$ \cite{yanaprb}.  Above $y = 1.2\%$, the carrier concentration begins to decrease, indicating a movement of the chemical potential away from $\mu_c$.  This is consistent with the donor nature of In dopants in lead telluride, as expected.  Between $y = 2.1 \%$ and $3.1\%$, the material has become n-type.  In the case that all Tl atoms are in the $+1$ valence state, one would expect full compensation of holes introduced by Tl dopants by an In concentration equal to the concentration of Tl dopants.  The fact that this occurs at a much higher concentration suggests either an increase in lead vacancies with increasing In concentration, which would add additional holes, or possibly even a mixed In valence.

Representative low temperature resistivity data are shown in figure \ref{rholowtintl}. For $y \leq 1.2\%$, a low-temperature resistance minimum at 8.5 K, similar to that observed in Pb$_{.99}$Tl$_{.01}$Te, is found. Increasing In concentrations lead to both a reduction in the residual resistivity and a suppression of the low-temperature minimum.  The magnitude of this resistance minimum as a function of $y$, defined as $\rho(1.8 K) - \rho(8.5 K)$ , is shown as a function of $y$ in panel E of figure \ref{trendsintl}.   The dependence of  the ratio of the resistance at room temperature to that at base temperature, known as the residual resistance ratio, or $RRR$, is shown in panel D of figure \ref{trendsintl}.

Heat capacity was measured between 0.35 K and 5 K on 5-12 mg single crystals using the relaxation method with a Quantum Design PPMS system equipped with a helium 3 cryostat.  Measurements were made in zero field and in an applied field of 1 T, to suppress the superconducting transition, at arbitrary orientations of the sample in the field. The electronic contribution, $\gamma$, was calculated from linear fits to $C / T$ vs T$^2$, for data taken below 1 K and in an applied field.

\begin{figure}
\centering
\includegraphics[width=3.5in]{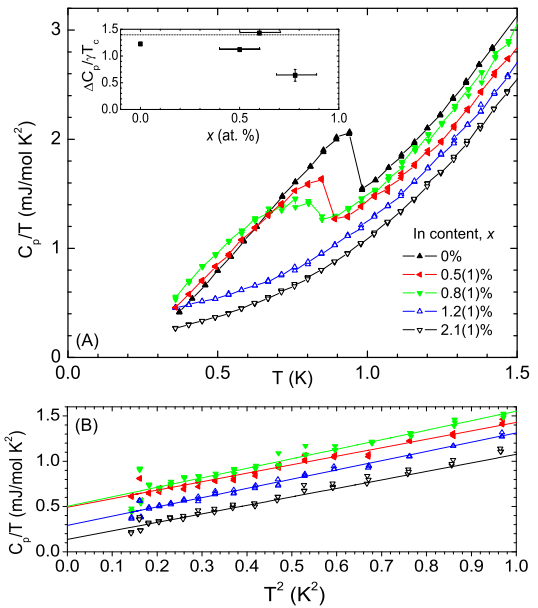}
\caption{\label{cpintl} (color online) (A) Specific heat divided by temperature of representative samples of Pb$_{.99-y}$Tl$_{.01}$In$_{y}$Te, with 0 $< y <$ 2.1\%.  Inset shows the magnitude of the specific heat jump at $T_c$ divided by the electronic coefficient to the specific heat, $\gamma$, and $T_c$.  The horizontal line in the inset represents the BCS weak coupling prediction of $\Delta C/\gamma T_c = 1.43$. (B) C$_p$/T as a function of T$^2$, for the same samples shown in panel (A).  Lines are linear fits to data taken below 1 K.}
\end{figure}

Heat capacity data at zero field are shown in figure \ref{cpintl} (A) for representative In concentrations.  Samples with In concentration $y < 0.8\%$ show sharp superconducting transitions, with $T_c$ decreasing with increasing $y$.  When $y = 0.8\%$, the superconducting transition is broad, and for $y > 0.8\%$, samples do not show a superconducting anomaly. The superconducting critical temperature, $T_c$, defined as the midpoint in the specific heat discontinuity, is shown as a function of In content, $y$, in panel B of figure \ref{trendsintl}.  Error bars were defined as $90\%$ to $10\%$ of the full transition height.  With increasing In counterdoping, $T_c$ is reduced from a value of 1.1 K for $y=0$ to less than 0.35 K by an In concentration of $y = 1.2\%$.

The electronic coefficient to the specific heat, $\gamma$, was derived from heat capacity data, taken in a field of 1 Tesla to suppress the superconducting transition.  Figure \ref{cpintl} (B) shows linear fits to $C/T$ as a function of $T^2$, for $T < 1$ K for representative In concentrations.  The electronic coefficient, $\gamma$, is shown as a function of $x$ in panel C of figure \ref{trendsintl}.  For $y < y_c$, Pb$_{.99-y}$Tl$_{.01}$In$_y$Te shows a constant value of  $\gamma = 0.505(5)$ mJ/molK$^2$, which is higher than the value of $0.4$ mJ/molK$^2$, expected for a Tl-- and In--free sample of similar carrier concentration, based on previous calculations using the known PbTe band structure \cite{yanaprb}. For $y > y_c$, where superconductivity is not observed, the materials show a reduced value of $\gamma <$ 0.2 mJ$/$mol K$^2$, much closer to the value expected for PbTe at those carrier concentrations \cite{yanaprb}.

All samples with $y < 1.2\%$ exhibit bulk superconductivity, with values of $\Delta C/\gamma T_c$ near the BCS value for weak coupling superconductors of $\Delta C/\gamma T_c =$1.43, shown in the inset to figure \ref{cpintl}.  Error bars were estimated by comparing linear extrapolations to $C/T$ over different temperature ranges above and below $T_c$.  The reduced value of $\Delta C/\gamma T_c$ for $y = 0.8\%$, as well as the broadened transition in this sample, likely reflects inhomogeneity in the material.  While there is no reason to expect a larger degree of inhomogeneity in material with this particular In concentration, such inhomogeneity would be more noticeable in a sample with a composition near a sharp transition to nonsuperconducting behavior. Resistivity data for two samples of In concentration, $y = 1.2\%$, were collected to a temperature of 50 mK in a dilution refrigerator, and were not found to superconduct. The particular case, $x = 1.2\%$, shows an intermediate value of $\gamma$ and $n_H$.  While we cannot rule out that the sample does superconduct with $T_c < 50$ mK, it is also possible that the effect of inhomogeneity for concentrations near $y_c$ results in a sample which shows a resistance minimum, but no bulk superconductivity, or that heating or current density effects have suppressed $T_c$.

The fact that electron microprobe analysis (EMPA) measurements found a wide standard deviation of the Tl content in all samples suggests inhomogeneity in the Tl concentration.  This is supported by the broadening of the superconducting anomaly in heat capacity data for $x = 0.8\%$  (figure \ref{cpintl}).  All materials were found to have a Tl concentration, $x_{min} > 0.7(4)\%$, suggesting that, in the absence of In counterdopants, they would all behave similar to Pb$_{1-x}$Tl$_x$Te, with $x > x_c$, as described in references \cite{yana} and \cite{yanaprb}.  The trends observed in figure \ref{trendsintl}, and discussed below, can therefore be attributed primarily to the changing In counterdopant concentration.

When $y$ increases beyond $y_c = 1.2\%$, the Hall number, $n_H$, is reduced, indicating that the chemical potential, $\mu$, has moved away from the critical value for inducing degeneracy between $+1$ and $+3$ Tl valence states, $\mu = \mu_c$ for In concentrations, $y > y_c$. Correlated with this change in $\mu$ is a suppression of $T_c$ to below 50 mK and  a suppression of the density of states at the Fermi level, $N(0)$, extracted from specific heat data.  This is similar to the effect of sodium and lithium counterdopants, described above \cite{kaidanov1987,chernik1982}. In addition, we also find an increase in the residual resistance ratio, $RRR$, and suppression of the low temperature resistance minimum (figure \ref{trendsintl}) as $y$ is increased beyond $y_c = 1.2\%$.  The observation that $adding$ additional In dopants actually $increases$ the $RRR$ is consistent with the hypothesis that that low temperature scattering in Pb$_{.99}$Tl$_{.01}$Te is dominated by anomalous scattering effects associated with the mixed-valence state, which is suppressed as the chemical potential moves away from $\mu_c$. The fact that the dependence of the parameters shown in figure \ref{trendsintl} on $y$ is slightly nonmonotonic, as shown in figure \ref{trendsintl}, is consistent with the variation in Tl content across the series and the inhomogeneity detected within each batch.  

In the case of lithium \cite{kaidanov1987} and sodium \cite{chernik1982, kaidanov1986} counterdopants, a correlation between elevated density of states, chemical potential pinning, and superconductivity was also observed.  The authors of those studies attributed this correlation to formation of a narrow Tl impurity band, in contrast to the charge-Kondo picture developed in references \cite{yana} and \cite{joerg}.  However, the additional correlation of Kondo-like scattering with superconductivity and chemical potential pinning noted in the present work cannot be explained in an impurity band model.

Additionally, recent angle-integrated photoemission spectra did not observe a peak in the denisty of states near the Fermi level in single crystals of Pb$_{.995}$Tl$_{.005}$Te at a temperature of 20 K \cite{nakayama2008}, well above the Kondo temperature of 6 K, reported in reference \cite{yana}.  The observation of elevated density of states when $\mu = \mu_c$ derived from low-temperature specific heat measurements in the present work suggests that this peak arises from a charge-Kondo resonance at low temperature, rather than a narrow impurity band, which would also be present at 20 K. 

In summary, the present work demonstrates a clear correlation between chemical potential pinning, superconductivity, and the presence of Kondo-like resistance upturn previously observed in this system \cite{yana,joerg}, independent of the Tl concentration.  This provides a distinction between spurious effects introduced by the dopant atoms themselves, such as those introduced by magnetic impurities, and effects resulting from the intrinsic mixed-valent state present when $\mu$ is near $\mu_c$, and supports the hypothesis that superconductivity in Pb$_{1-x}$Tl$_x$Te arises from a negative U mechanism, which also induces a charge-Kondo effect.

The authors acknowledge the contribution of Yana Matsushita to data taken for In-free material.  We also thank Robert E Jones for technical assistance in EMPA measurements.  This work is supported by the Department of Energy, Office of Basic Energy Sciences under contract DE-AC02-76SF00515.

\end{document}